\documentclass[epjCONF,columns]{svjour} 
\usepackage{graphics}
\usepackage[varg]{txfonts} 
\usepackage[latin1]{inputenc}
\session-title{Hadron Collidel Physics Symposium HCP2011}
\begin{document}
\hyphenation{mo-no-chro-ma-tic se-arch-ed back-ground}
\title{Search for Physics Beyond the Standard Model at BaBar and Belle}
\author{G.Calderini\inst{1}\inst{2}\fnmsep\thanks{\email{giovanni.calderini@lpnhe.in2p3.fr}} \thanks{on behalf of BaBar and Belle Collaborations}}
\institute{Laboratoire de Physique Nucleaire et des Hautes Energies LPNHE Paris 
\and Universit\'a di Pisa and Istituto Nazionale di Fisica Nucleare, Sez. di Pisa}
\abstract{
Recent results on the search for new physics at BaBar and Belle B-factories are presented. The search for a light Higgs boson produced in the decay of different $\Upsilon$ resonances is shown. In addition, recent measurements aimed to discover invisible final states produced by new physics mechanisms beyond the standard model are presented.      
} 
\maketitle
\section{Introduction}
\label{intro}
A light Higgs boson is foreseen in many extensions of the Standard Model. 
In the limit of ($m_H < 2m_b$) it may become accessible through $\Upsilon$ 
resonances \cite{hiller,gunion}. 
Under this scenario, B-factories represent an ideal discovery environment, 
and they complete the existing results from high energy electron-positron machines 
as LEP or more recently coming from experiments at hadron machines as 
Tevatron and LHC. 

\section{BaBar and Belle}
\label{sec:1}
\begin{figure}[h]
\resizebox{0.95\columnwidth}{!}{\includegraphics{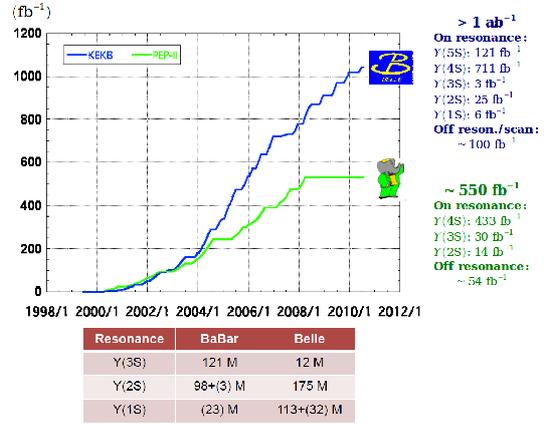} }
\caption{Evolution of integrated luminosity as a function of time for the Belle and BaBar experiments with the breakdown of data taken at the different energies. The table at the bottom represents in particular the number of millions of events collected by the two experiments at the three resonances used for the search reviewed in this paper. Numbers in parentheses are the feed-down contributions from higher resonances.}
\label{fig:1}
\end{figure}

%
Tha BaBar Collaboration at PEP-II (SLAC) \cite{babar} and the Belle Collaboration at KEKB (Tsukuba) 
\cite{belle} have been successfully taking data since 1999 mainly around and at the energy of the 
$\Upsilon(4s)$ resonance. In the last part of their respective physics programmes, the center-of-mass 
energy has been varied enough to study other $\Upsilon$ resonances. The total accumulated data has 
been of more of $1~ab^{-1}$ for Belle and about $550~fb^{-1}$ for BaBar.
A picture showing the integrated luminosity of the two experiments with the breakdown at the different 
energies is shown in Fig.\ref{fig:1}. The table at the bottom of the picture presents the number of 
millions of events collected at the three resonances used in the searches reviewed in this paper. 
The numbers in parentheses are the additional events coming from the feed-down contributions from 
higher-energy resonances. 

\section{Direct searches for a light CP-odd Higgs}
\label{sec:2}
\subsection{\boldmath{$\Upsilon(2s,3s) \rightarrow \gamma A^0 \rightarrow \gamma \mu^+ \mu^-$}}
\begin{figure}
\begin{center}
\resizebox{0.75\columnwidth}{!}{\includegraphics{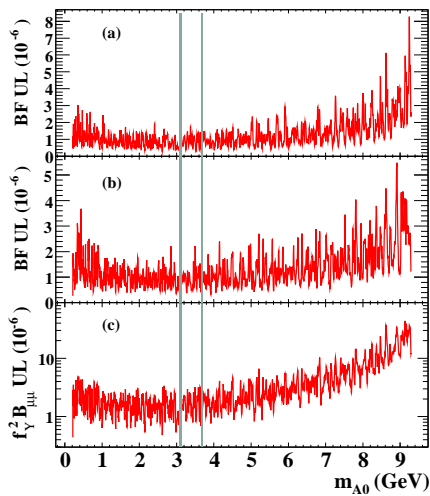} }
\end{center}
\caption{The $90\%$ CL upper limit as a function of the $A^0$ mass for the $\Upsilon \rightarrow \gamma A^0(\mu\mu)$ extracted from the $\Upsilon(2s)$ (plot a) and
$\Upsilon(3s)$ samples (b), and the limit on the product of the branching fraction by the effective coupling $f_Y^2$ (c). The shadowed regions correspond to the $J/\psi$ and $\psi(2s)$ resonances which are excluded from the search.}
\label{fig:babar:up23gmumu}
\end{figure}

This analysis performed by BaBar \cite{babar:up23gmumu} is based on the selection of a photon with a minimum center-of-mass energy $E^*_\gamma > 0.2 GeV$ and two oppositely charged tracks with a vertex compatible with the luminous 
region. A muon mass hypothesis is assigned to the two tracks and after a kinematic fit to the $\gamma\mu\mu$ candidate, the signal is searched in the form of a peak in the reduced mass distribution 
$m_R=\sqrt{m_{\mu\mu}^2-4m_\mu^2}$.
No significant excess of events above the background was observed in both 
the $\Upsilon(2s)$ and the $\Upsilon(3s)$ samples and $90\%$ confidence 
level (CL) limits are established: 

\noindent
$B(\Upsilon(2s) \rightarrow \gamma A^0(\mu^+\mu^-)<(0.26 \div 8.3)10^{-6}$ and $B(\Upsilon(3s)\rightarrow\gamma A^0(\mu^+\mu^-)<(0.27-5.5)10^{-6}$ for a mass hypothesis in the range $0.212 \leq m_{A^0} \leq 9.3~GeV$. Upper limits as a function of the $A^0$ mass are shown in Fig. \ref{fig:babar:up23gmumu}.  

\subsection{\boldmath{$\Upsilon(3s) \rightarrow \gamma A^0 \rightarrow \gamma \tau^+ \tau^-$}}
\begin{figure}[h]
\begin{center}
\resizebox{0.8\columnwidth}{!}{\includegraphics{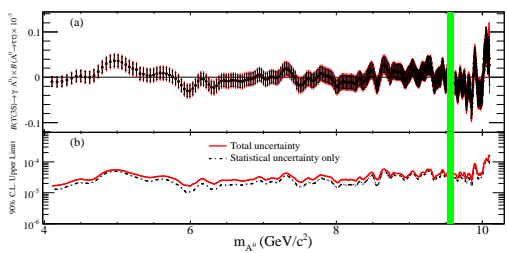} }
\end{center}
\caption{The product branching fraction and the $90\%$ CL upper limit as a function of the $A^0$ mass for the $\Upsilon(3s) \rightarrow \gamma A^0$ with the $A^0 \rightarrow \tau^+\tau^-$ analysis (BaBar).} 
\label{fig:babar:up3gtt}
\end{figure}

BaBar has also searched for an $A^0$ signal in the decay of 
$\Upsilon(3s)$ $\rightarrow \gamma \tau^+ \tau^- $ \cite{babar:up3gtt}. 
The two tau candidates are reconstructed in the leptonic channel, and events with two tracks 
identified as $e$ or $\mu$ ($ee$, $e\mu$ and $\mu\mu$ combinations) and a photon with 
$E_\gamma > 0.1~GeV$ are selected. The dominant background is represented by QED radiative tau 
pairs $e^+e^-\rightarrow \tau^+\tau^-\gamma$. Since the event is not fully reconstructed, the 
signature of signal is given by a peak in the $E_\gamma$ distribution, which is scanned in the 
range $4.03 < m_{A^0} < 10.10$ GeV after peaking background removal. No significant signal is 
found and an upper limit is set 
$B(\Upsilon(3s) \rightarrow \gamma A^0(\tau^+\tau^-)<(1.5 \div 16)10^{-5}$ at 90 \% confidence level, 
as shown in fig. \ref{fig:babar:up3gtt}.     

\subsection{\boldmath{$\Upsilon(2,3s) \rightarrow \gamma A^0$} with 
\boldmath{$A^0 \rightarrow$} hadrons}
\begin{figure}[h]
\begin{center}
\resizebox{0.8\columnwidth}{!}{\includegraphics{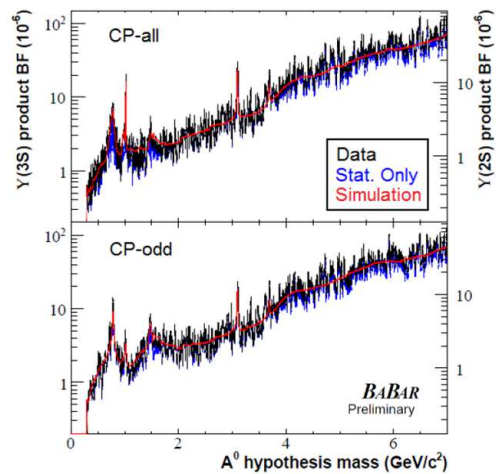} }
\end{center}
\caption{The product branching fraction as a function of the $A^0$ mass for the $\Upsilon(2s,3s) \rightarrow \gamma A^0$ with $A^0 \rightarrow$ hadrons analysis (BaBar).} 
\label{fig:babar:up23ghad}
\end{figure}

A BaBar recent analysis \cite{babar:up-had} involves the hadronic decay mode of the $A^0$. 
In this case the event can be fully reconstructed. The highest-energy photon in the event 
($E_{\gamma} > 2.2(2.5)$ GeV in the $\Upsilon(2s)$ and $\Upsilon(3s)$ selections) after a 
$\pi^0$ and $\eta$ veto is chosen. The sum of all 4-momenta of the remaining objects 
($K_s, K, \pi, p, \pi^0$ and leftover $\gamma$) is taken as the $A^0$ candidate. Invariant mass 
distributions are scanned for peaks in the $\Upsilon(2s)$ and $\Upsilon(3s)$ selections. The
priduct branching fractions observed are shown if fig. \ref{fig:babar:up23ghad}.

No significant excess of events is observed and a limit is set 
$B(\Upsilon(ns)$ $\rightarrow \gamma A^0($hadrons$))<(0.1 \div 8)10^{-5}$ at 90\% confidence level.   

\subsection{More recent results involving \boldmath{$\Upsilon(1s) \rightarrow \gamma A^0$} 
with the $A^0$ decaying into $\mu^+\mu^-, \tau^+\tau^-$ or hadrons$)$}
More recent results are worth to be mentioned, involving the $\Upsilon(1s)$ transition to $\gamma A^0$ with $A^0$ reconstructed to visible states. In particular Belle looks at the 
$\Upsilon(1s) \rightarrow \gamma A^0$ with the $A^0 \rightarrow \tau^+\tau^-$ \cite{belle:up1tt}. 
The analysis is still underway and, as in the previous examples of reconstruction of $A^0$ into 
$\tau$ pairs, is based on a study of the $E_{\gamma}$ distribution. Taus are presently reconstructed 
in the leptonic channel $e\mu$, while $ee$ and $\mu\mu$ combinations will be added soon. In addition, 
BaBar is studying the $\Upsilon(1s) \rightarrow \gamma A^0$ with $A^0 \rightarrow \mu^+\mu^-, \tau^+\tau^-$ or hadrons $)$ using a di-pion tag to identify the $\Upsilon(1s)$ from the 
$\Upsilon(3s) \rightarrow \pi^+\pi^- \Upsilon(1s)$ transition. 
   
\subsection{Invisible decays}
In some nMSSM models with $\chi$ as Lightest Supersymmetric Particle (LSP), the dominant decay mode of the $A^0$ could be $A^0 \rightarrow \chi^0 \overline{\chi}^0$.
For this reason, all the analyses involving invisible decays of the $A^0$ have a special interest, given 
also the implications for Dark Matter existence.  
\subsubsection{\boldmath{$\Upsilon(3s) \rightarrow \gamma A^0$} with 
\boldmath{$A^0 \rightarrow$} invisible states.}
This BaBar analysis \cite{babar:inv:3s} is based on a search for a monochromatic photon in the event in conjunction with missing energy. A peak in the center-of-mass $E^*_{\gamma}$ distribution is searched and the invariant mass of the recoil system is calculated. A scan to identify an excess of events is performed. No significant signal is observed and also in this case an upper limit is set
$B(\Upsilon(3s) \rightarrow \gamma A^0($invisible$))<(0.7 \div 31)10^{-6}$ at 90\% confidence level. 

\subsubsection{\boldmath{$\Upsilon(1s) \rightarrow \gamma A^0$} with 
\boldmath{$A^0 \rightarrow$} invisible states}. 
The case of $\Upsilon(1s)$ decay involving invisible products is of special interest. In fact, the SM 
process $\Upsilon(1s) \rightarrow \gamma\nu\overline{\nu}$ is not observable at the present experimental 
sensitivity (B $\approx 10^{-5}$) \cite{yeghiyan}. 
At the same time the branching fraction of $\Upsilon(1s) \rightarrow \gamma A^0$ could be as large 
as $5\times10^{-4}$ depending on the mass of the $A^0$ and the couplings \cite{gunion}. 
An observation of $\Upsilon(1s)$ decays with significant missing energy could be a sign of new physics. 

\begin{figure}[h]
\begin{center}
\resizebox{1.1\columnwidth}{!}{\includegraphics{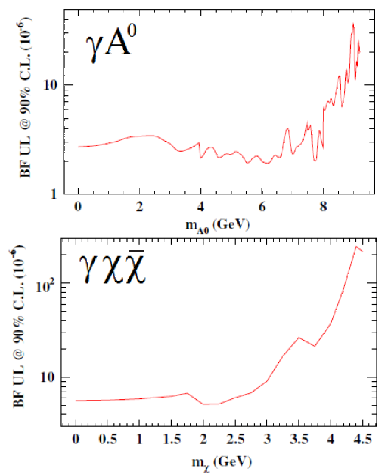} }
\end{center}
\caption{The $90\%$ CL upper limit as a function of the $A^0$ mass and $\chi$ mass for the $\Upsilon \rightarrow \gamma A^0$ and $\Upsilon \rightarrow \gamma \chi\overline{\chi}$ searches.} 
\label{fig:babar:up2ppup1}
\end{figure}

In a BaBar analysis \cite{babar:up3ppup1}, the $\Upsilon(1s)$ is tagged from the $\Upsilon(3s)\rightarrow \pi^+\pi^- \Upsilon(1s)$ transition. Both the resonant two-body decay $\Upsilon(1s)\rightarrow \gamma A^0$ and the non-resonant three-body decay $\Upsilon(1s)\rightarrow \gamma \chi\overline{\chi}$ are analyzed. 
Two pions of opposite charge and a single energetic photon with $E^*_{\gamma} \le 0.15$ GeV plus a large amount of missing energy and momentum are required in the event. Sources of background as the 
$\Upsilon(1s)\rightarrow \gamma K_LK_L$ and $\Upsilon(1s)\rightarrow \gamma n\overline{n}$ are reduced by using a hadron calorimeter (IFR) based veto. The signal yield is extracted as a function of 
$m_{A^0}(m_\chi)$ in the interval $0 \leq m_{A^0} \leq 9.2$ GeV ($0 \leq m_{\chi} \leq 4.5$ GeV), 
using two kinematic variables: the di-pion recoil mass $M_{rec}$ and the missing mass squared $M_X^2$. 
No significant excess of events above the background is observed and upper limits are set 
at 90\% confidence level on 
$B(\Upsilon(1s) \rightarrow \gamma A^0($invisible$)<(1.9 \div 37)10^{-6}$ and 
$B(\Upsilon(1s) \rightarrow \gamma \chi\overline{\chi})<(0.5 \div 24)10^{-5}$ 
(Fig.\ref{fig:babar:up2ppup1}). 

\subsubsection{Direct decay of \boldmath{$\Upsilon(1s)\rightarrow$} invisible states.}
\begin{figure}[h]
\begin{center}
\resizebox{\columnwidth}{!}{\includegraphics{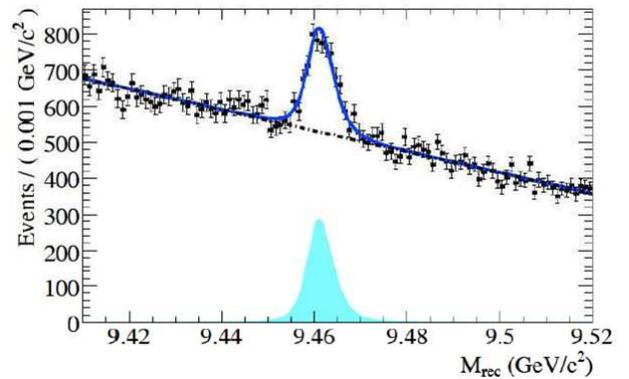} }
\end{center}
\caption{The $\pi\pi$ recoil mass distribution $m_{rec}$ in the $\Upsilon(1s) \rightarrow$ invisible 
anaysis. Data are shown as points, while the overall fit (solid) and the combinatoric contribution
(dashed) are shown. A large excess of events is still observed (shadowed) but it is consistent with the
peaking background expected contribution.} 
\label{fig:babar:up3ppup1}
\end{figure}

BaBar also searched for a Dark Matter candidate in the direct decay of $\Upsilon(1s)$ to invisible 
states \cite{babar:up3ppup1.2}. Also in this case, the $\Upsilon(1s)$ is tagged with the help of the 
di-pion system in the $\Upsilon(3s)\rightarrow \pi^+\pi^-\Upsilon(1s)$ transition. Two oppositely charged 
pions with no other detector activity in the event are required. The mass of the recoil system is 
reconstructed. In addition to the combinatoric events, there are other important sources of peaking 
background. In particular the $\Upsilon(3s) \rightarrow \pi^+\pi^-\Upsilon(1s)$ decay, where the 
$\Upsilon(1s)$ final state particles (mainly lepton pairs, low energy particles or other 
non-interacting neutral hadrons) are undetected. The contribution from these peaking sources is 
estimated from simulation and validated on data in a control sample with similar requirements 
as the signal sample. After the combinatoric background subtraction, a still significant signal eccess
of $2326\pm105$ events is observed (see Fig.\ref{fig:babar:up3ppup1}). At the same time the peaking
background expected contribution is determined to be of $2444\pm123$ events, which is fully consistent
with the excess found.

After total background subtraction a signal yield consistent with zero ($-118 \pm 105 \pm 24$) is found 
in the expected region and an upper limit of 
$B(\Upsilon(1s)\rightarrow $invisible$) < 3.0 \times 10^{-4}$ at the 90\% confidence level is obtained.   

\subsection{Conclusions}
The searches for a light Higgs boson $A^0$ by BaBar and Belle experiments using the datasets collected at the $\Upsilon$ resonances are presented.No evidence is found, which strongly contraints the available parameter space of the nMSSM. In addition the result of the search for invisible decays of the 
$\Upsilon(1s)$ is reported, to look for light Dark Matter candidates.  

%
%

\end{document}